\documentclass{aa}
\usepackage{graphicx,natbib,url}
\usepackage{color}
\usepackage{amsfonts}
\usepackage{amssymb}
\bibpunct{(}{)}{;}{a}{}{,} 
\usepackage{txfonts}
\usepackage{xspace}
\newcommand{\ie}{i.e.,~}
\newcommand{\cf}{cf.,~}

\newcommand{\text}[1]{\rm  #1}

\newcommand{\urlfn}[1]{\footnote{\url{#1}}}

\newcommand{\MS}{\text{magnetosphere}\xspace}
\newcommand{\emep}{\text{e$^+$/e$^-$}\xspace}

\newcommand{\secs}{{\rm  s}}

\begin{document}
\title{Fast radio bursts: the last sign of supramassive neutron stars} 
\author{Heino Falcke\inst{1,2,3} \and  Luciano {Rezzolla}\inst{4,5}}

\institute{Department of Astrophysics, Institute for Mathematics,
  Astrophysics and Particle Physics, Radboud University Nijmegen,
  P.O. Box 9010, 6500 GL Nijmegen, The Netherlands
\and 
ASTRON, Oude Hoogeveensedijk 4, 7991 PD Dwingeloo, The Netherlands \and
Max-Planck-Institut f\"ur Radioastronomie, Auf dem H\"ugel 69, 53121
Bonn, Germany
\and 
Max-Planck-Institut f\"ur Gravitationsphysik, Albert-Einstein-Institut,
Potsdam, D-14476, Germany
\and 
Institut f\"ur Theoretische Physik, Frankfurt am Main, D-60438, Germany}

\date{Submitted: May 30, 2013, accepted: Jan 20, 2014} 

\abstract
{Several fast radio bursts have been discovered recently, showing a
  bright, highly dispersed millisecond radio pulse. The pulses do not
  repeat and are not associated with a known pulsar or gamma-ray
  burst. The high dispersion suggests sources at cosmological
  distances, hence implying an extremely high radio luminosity, far
  larger than the power of single pulses from a pulsar.}
{We suggest that a fast radio burst represents the final signal of a
  supramassive rotating neutron star that collapses to a black hole due
  to magnetic braking. The neutron star
  is initially above the critical mass for non-rotating models and is
  supported by rapid rotation. As magnetic braking constantly reduces the
  spin, the neutron star will suddenly collapse to a black hole 
  several thousand to million years after its birth.}
  {We discuss several formation scenarios for supramassive neutron stars
    and estimate the possible observational signatures {making use of
    the results of recent numerical general-relativistic calculations.}}
  {While the collapse will hide the stellar surface behind an event
    horizon, the magnetic-field lines will snap violently. This
    can turn an almost ordinary pulsar into a bright radio
    ``blitzar'': Accelerated electrons from the travelling magnetic
    shock dissipate a significant fraction of the magnetosphere and
    produce a massive radio burst that is observable out to
    $z>0.7$. Only a few percent of the neutron stars needs to be
    supramassive in order to explain the observed rate.}
  {We suggest the intriguing possibility that fast radio bursts might
    trace the solitary and almost silent formation of stellar mass black
    holes at high redshifts. These bursts {could be an electromagnetic
    complement to gravitational-wave emission} and reveal a new
    formation and evolutionary channel for black holes and neutron stars
    that are not seen as gamma-ray bursts. If supramassive neutron stars
    are formed at birth and not by accretion, radio observations of these
    bursts could trace the core-collapse supernova rate throughout the
    universe.}

\authorrunning{Falcke \& Rezzolla}
\titlerunning{Fast radio bursts and collapse of supramassive star}

\maketitle

\section{Introduction}

\label{s:intro}
{Recently, a number of isolated fast radio bursts (FRBs) have
  been discovered that last for only a millisecond and which may come
  from cosmological distances
  \citep{LorimerBailesMcLaughlin2007a,ThorntonStappersBailes2013a}. What
  could possibly produce such a bright emission in such a short time?
  A natural explanation may be gravitational collapse involving
  neutron stars (NSs) or stellar mass black holes (BHs).  } Typically,
  the formation of NSs and BHs are associated with some rather
  energetic observational signatures across all wavelengths, such as a
  supernova (SN) or a gamma-ray burst (GRBs).  The latter are
  short-term flares of X-ray and gamma-ray emission, lasting only a
  fraction of seconds to {tens of seconds, sometimes with an
    extended afterglow}. The total energy radiated in a GRB is $\sim
  10^{48-50}$ erg \secs{}$^{-1}$ and the bright emission {was
    explained initially} in a fireball model
  \citep{CavalloRees1978a,Paczynski1986a,Eichler89}, where a
  significant fraction of the energy is thermalized in an optically
  thick outflow eventually radiated in the form of high-energy
  emission \citep[see][for recent reviews]{Nakar:2007yr,Lee:2007js}.

{Short GRBs, of duration less than 2 s, are thought to be
  associated with NS-NS mergers and not to trace well star formation
  \citep{GehrelsSarazinOBrien2005a}. Their average timescale is around
  50 ms \citep{Nakar:2007yr} with some spread.} Long GRBs, with a
duration longer than 2 s, may be associated with the SN of a massive
star, thereby well tracing cosmic star formation
\citep{WoosleyBloom2006a}. For the latter scenario, the GRB is
suggested to be due to a plasma jet that propagates through the dense
outer layers of the exploding star \citep{Woosley1993a}. Hence, baryon
loading and particle acceleration in internal or external shocks play
an important role in the observational appearance of GRBs. However, do
all forms of collapse lead to such bright observational signatures?

We here discuss the collapse of an isolated and magnetized
supramassive rotating neutron star (SURON) to a BH in a rarefied
environment. Such a collapse would be inevitable if a rapidly spinning
NS was formed above the critical mass for a non-rotating NS. Over
time, magnetic braking would clear out the immediate environment of
the star and slow it down. With a significant delay after the creation
of the neutron star in a supernova implosion, the SURON will eventually
collapse almost instantly. The formation of an event horizon over the
free-fall timescale, i.e.~$<$ 1 ms, during the collapse would
immediately hide most of the matter and radiation apart from the
SURON's magnetosphere. Instead, the \MS will experience a violent
disruption leading to a strong magnetic shock wave travelling outwards
near the speed of light and producing radio emission. Hence, the
observational signatures of such a system would be quite different
from those of short or long GRBs and more akin to that of pulsar
emission.

{The collapse of a SURON near break-up spin has already been
  considered by \citet{VietriStella1998a}, who suggested that, while
  the collapse itself is ``silent'', accretion from a torus,
  consisting of the mass shed from the stellar equator, onto the newly
  formed black hole could power a wind that is detectable as an
  afterglow. The ``silence'' in a NS collapse comes from the fact that
  the $pdV$-work done by the contracting star which could heat it up,
  cannot escape via conduction or neutrinos because of the very short
  timescales involved and thus ends up in the BH.}

{We now know, through general-relativistic simulations of the collapse of
  SURONs near break-up, that such a torus does not seem to form [see
    Fig. 16 of~\citet{BaiottiHawkeMontero2005a}]. At the same time,
  recent general-relativistic simulations in resistive
  magnetohydrodynamic (MHD) also tell us that the collapse of magnetized
  NSs is accompanied by a strong electromagnetic
  emission~\citep{LehnerPalenzuelaLiebling2012a,
    DionysopoulouAlicPalenzuela2013a}. Hence, a collapsing SURON is not entirely
  ``silent'' and, as we will discuss, by simply using the power stored in
  the magnetosphere, a detectable prompt signature should be produced in
  the radio regime.}

In fact, the recently discovered fast radio bursts (FRBs) may make this
scenario an intriguing possibility. In 2007 Lorimer et
al. \nocite{LorimerBailesMcLaughlin2007a} discovered a single bright and
highly dispersed radio flash in archival pulsar survey data of the Parkes
telescope that was not associated with any known pulsar or GRB. In the
meantime, more of these events have been found
\citep{KeaneStappersKramer2012a,ThorntonStappersBailes2013a}. While a
terrestrial origin of these radio signals can not be fully excluded
\citep{Burke-SpolaorBailesEkers2011a} and the bursts have not been
confirmed by other telescopes yet
\citep[e.g.,][]{BowerWhysongBlair2011a,SiemionBowerFoster2012a}, an
origin of these one-off radio bursts at cosmological distances now seems
a viable interpretation
\citep{ThorntonStappersBailes2013a,LorimerKarastergiouMcLaughlin2013a}.

{Theoretical interpretations of the Lorimer burst have been rare
  until now and no canonical picture has emerged yet
  \citep{ThorntonStappersBailes2013a}. A connection with short hard GRBs
  and merging NSs has been proposed
  \citep{PshirkovPostnov2010a,Lyutikov2013a}, as well as supernova
  explosions in a binary system impacting a NS magnetosphere
  \citep{EgorovPostnov2009a}, reconnection in the magnetosphere of
  neutron stars \citep{Somov2011a}, or even superconducting cosmic
  strings \citep{CaiSabancilarSteer2012a} and the evaporation of BHs in
  the presence of extra spatial dimensions
  \citep{KavicSimonettiCutchin2008a}. In light of the new observations it
  may be worth revisiting these scenarios.}\footnote{{Since the
    submission of this paper, other interpretations and models have been
    put forth to explain FRBs; see, e.g., \citet{Totani2013a, Katz2013a,
      KashiyamaIokaMeszaros2013a, LoebShvartzvaldMaoz2014a, Luan2014a}.}}

The basic properties of the six currently known radio bursts can be
summarized as follows: they are short with timescales $\Delta t \la 1$
ms; the radio fluxes are typically around 1 Jy at GHz frequencies,
with the brightest and closest one reaching over 30 Jy; no associated
GRBs have been seen at the time of the bursts; the rate is about
$0.25$ per square degree per day; the radio burst is dispersed, with
shorter wavelengths preceding longer ones, following a $\Delta t
\propto \lambda^{2}$ law. The dispersion is similar to what is seen in
pulses from Galactic pulsars due to free electrons in the interstellar
medium. However, the derived dispersion measures (DM), in the range of
a few hundred to thousand pc cm$^{-2}$, far exceed the Galactic DM in
those directions. Hence, dispersion due to the intergalactic medium
has been suggested, {which provides distance estimates of several
  Gpc and redshifts in the range $z=0.3-1$. In two cases, there is
  also evidence for a frequency-dependent scattering tail as expected
  for bursts passing through a turbulent ionized medium
  \citep{LorimerBailesMcLaughlin2007a,ThorntonStappersBailes2013a}.}

The distribution of the few bursts would be consistent with the
cosmological star formation or core-collapse SN rate. The apparent
isotropic luminosity of these bursts {at a mean redshift of $z=0.7$
  and luminosity distance $D_{\rm l}\sim4.3$ Gpc} (using $H_0=79.4$ km
\secs{}$^{-1}$ Mpc$^{-1}$, $\Omega_{\rm m}=0.27$, $\Omega_{\rm vac}=0.73$),
for an observed spectrum $S_\nu\propto\nu^\alpha$, would be
\begin{equation}\label{eq:frbluminosity}
L = 3 \times 10^{43} 
\left({{\nu}\over1.4\,{\rm GHz}}\right)^{1+\alpha}
\left({{S_\nu}\over1\,{\rm Jy}}\right)
\left({D_{\rm l}\over11\,{\rm Gpc}}\right)^2\; {\rm erg\,\secs{}^{-1}}\,.
\end{equation}
This luminosity is more than nine orders of magnitude brighter than a
giant kJy flare from the Crab pulsar. On the other hand, integrated
over 1 ms this yields $3 \times 10^{40}$ erg, which is only a tiny
fraction of a SN energy and much less than a typical GRB. Of course,
the total luminosity could be somewhat higher if the spectrum were
flat ($\alpha\simeq0$) and would extend to higher frequencies.

{Finally, we point out that the observed timescale of $\la1$~ms,
  together with the high luminosity and non-repetitive nature is very
  constraining. This is much shorter than that of SNe, long and short
  GRBs, or of merging NS binaries. This points towards the shortest
  timescale available for compact objects, namely, the free-fall
  timescale of NSs. Also, the lack of GRB signatures and the
  appearance of high-brightness temperature emission, implying
  coherent emission processes, is very reminiscent of emission
  mechanisms of pulsars. Hence, in the following we propose a scenario
  that is able to address all these issues.}

\section{Braking-induced collapse of a supramassive NS} \label{sec:formation}

The basic scenario we consider here assumes an initial state of a
magnetized NS with (gravitational) mass $M$ and spin frequency
$\Omega=2\pi\,\tau^{-1}$, where $M_{\rm max}(\Omega)<M<M_{\rm
  max}(\Omega_{\rm K})$ and $\tau$ is the spin period. Here $M_{\rm
  max}(\Omega)$ is the maximum mass above which a NS is unstable to a
collapse to a BH and $\Omega_{\rm K}$ is the maximum spin a NS can have,
sometimes called Keplerian or break-up spin. In the absence of rotation,
$M_{\rm max}(0)$ would be the equivalent of the Chandrasekhar limit for
NSs. Its exact value depends on the still unknown equation of state (EOS)
of nuclear matter, but NSs with masses larger than $M_{\rm max}(0)$ can
still be stable if they are supported by centrifugal forces. If we assume
the NS is spinning at a fraction $f$ of the break-up spin, \ie $\Omega =
f \Omega_{\rm K}$, then the resulting period is
{
\begin{equation}\label{eq:period}
\tau_{\rm rot} = 2 \pi \,f^{-1}\, \sqrt{\frac{R^3}{G M}} 
\simeq 3.8 \; f^{-1}_{0.1} \, r_{10}^{3/2}\, m_2^{-1/2} \;\;\text{ms}\,,
\end{equation}
}\noindent
where $M = m_2\, 2.3 M_{\odot}$ is the mass of the NS, $R=r_{10}\,
10\,{\rm km}$ its radius, $G$ the gravitational constant, {and
  $f_{0.1}=0.1\,f$}. Standard magnetic dipole radiation would then lead to a
braking of the system on a timescale
{
\begin{equation}
\label{eq:tbrake}
\tau _{\text{braking}} =  \frac{6 c^3 
M \sin^{-2} (\alpha_{\rm B})}{5 B^2 R^4 \Omega ^2} \simeq
3.1\, f^{-2}_{0.1} \,r_{10}^{-1}b_{12}^{-2}\;\;\text{kyr}\,,
\end{equation}
}\noindent
where $B=b_{12}\, 10^{12}\, {\rm G}$ is the magnetic field, $c$ is the
speed of light, and $\alpha_{\rm B}\simeq45^\circ$ is the magnetic
pitch angle. Hence, similar to highly magnetized pulsars
\citep{DuncanThompson1992a} the source, if highly spinning, would slow
down significantly within a few hundred to a few thousand years. If
the NS is supramassive it will collapse to a BH and hence disappear
from the general pulsar population.

Calculating the stability of rotating NSs is far from trivial. We here
make use of fully general-relativistic numerical calculations where the
star is modeled as a uniformly rotating polytrope with index
$\Gamma=2$~\citep{TakamiRezzollaYoshida2011a} and a polytropic constant
chosen so that $M_{\rm max}(0)=2.1 M_\odot$ to match recent observations
of high-mass NSs~\citep{AntoniadisFreireWex2013a}. Although simplified,
our EOS provides here a simple and overall realistic reference.

\begin{figure}
\resizebox{\hsize}{!}{\includegraphics{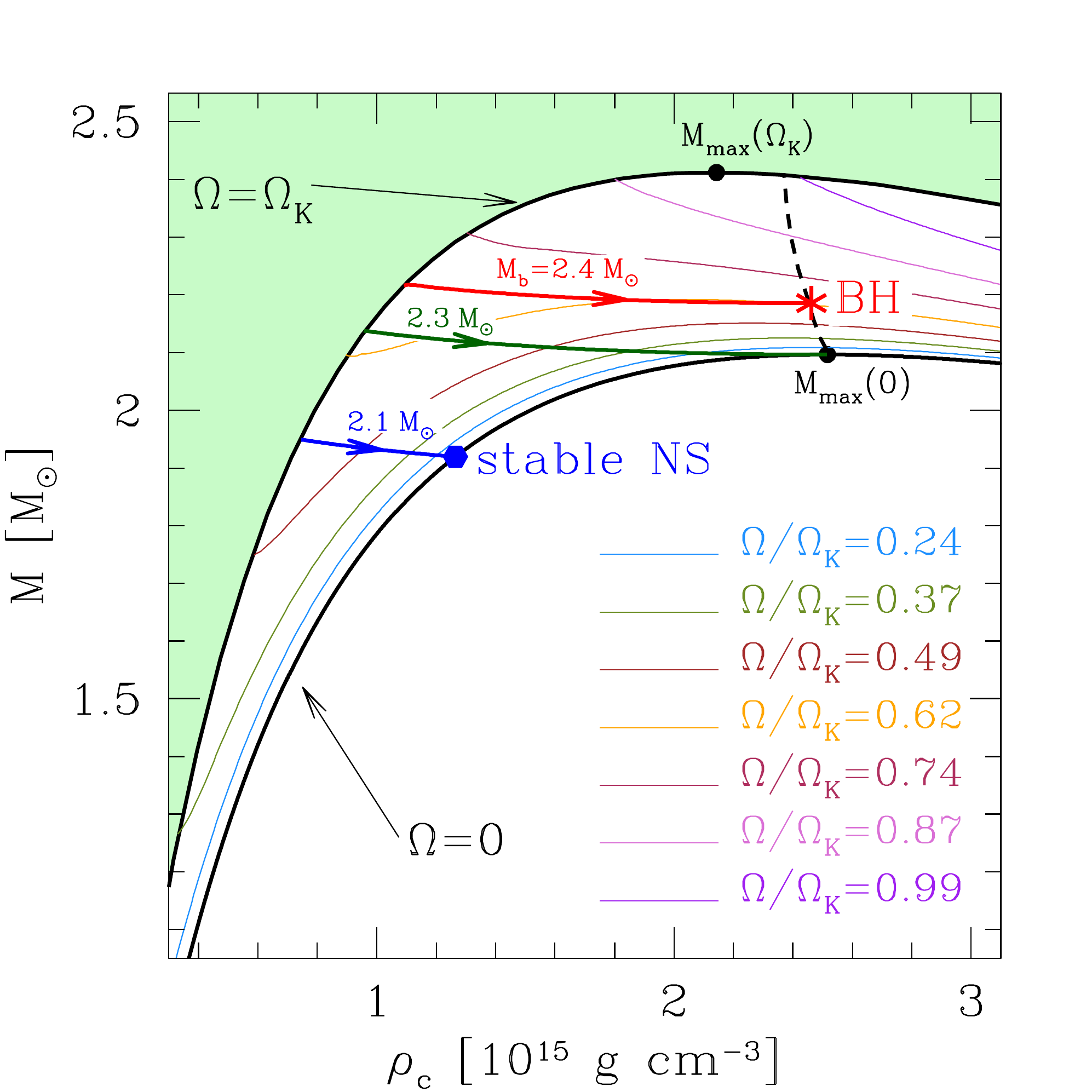}}
\caption{\label{fig:massradius}Gravitational mass versus central
  rest-mass density for NS models with different spins. The green shaded
  area is the region where no equilibrium models are possible because of
  excessive spin. The black solid lines are sequences of models that are
  non-rotating, $\Omega=0$, or rotating at break-up, $\Omega=\Omega_{\rm
    K}$, respectively (see text for more details). The arrows indicate
  example tracks of NSs with baryonic masses of 2.1, 2.3, and 2.4
  $M_\odot$ as they slow down due to magnetic braking.}
\end{figure}

Figure \ref{fig:massradius} illustrates the results of these calculations
in a diagram showing the \textit{gravitational} mass $M$, versus the
central rest-mass density $\rho_c$. The black solid lines are sequences
of models that are non-rotating, $\Omega=0$, or rotating at break-up,
$\Omega=\Omega_{\rm K}$, respectively. Hence, the green shaded area
indicates the region where no equilibrium models are possible because
they would be past the break-up limit. Shown as coloured lines are
sequences of constant angular frequency normalized to the maximum
possible spin frequency, \ie $\Omega/\Omega_{\rm K}$. We also show three
sequences of constant \textit{baryon} mass $M_b$, which can therefore be
interpreted as evolutionary tracks of a NS as it spins down during its
life. It is then easy to realize that, for instance, a SURON with
$M_b=2.4\,M_{\odot}$ (red solid line) could be produced near the break-up
limit (but also at smaller rotation rates) and then spin down while
maintaining its baryon mass. This corresponds to a motion to the right in
Fig.~\ref{fig:massradius}, during which the NS reduces its gravitational
mass (it contracts because of the decreased spin). This motion terminates
around the stability line, the locus of the maxima of sequences of
constant angular momentum, beyond which the star collapses to a Kerr BH
(shown as a black dashed line).  Similarly, a SURON with
$M_b=2.3\,M_{\odot}$ (green solid line) would also spin down, but now to
a zero spin frequency, when it reaches $M_{\rm max}(0)$. At that point it
will have become a spherical star and any perturbation will
induce its collapse to a Schwarzschild BH. Finally, a normal NS with
$M_b=2.1\,M_{\odot}$ (blue solid line) would also spin down to a zero
frequency, but end up on the stable branch of equilibrium models,
remaining there ever after.

In essence, for our representative EOS {\it any} spinning NS with a $M >
2.1 \, M_\odot$ ($M_b > 2.3 \, M_\odot$), will eventually collapse to a
BH. The {lifetime of the SURON will be of the order $\sim \tau_{\rm
    braking}$} and since this depends on $f^{-2}$, it follows from
Fig.~\ref{fig:massradius} that NSs which are just a few percent above
$M_{\rm max}(0)$ have the longest lifetimes and can collapse millions of
years after their formation. On the other hand, a NS with $M_b \simeq
2.4\, M_\odot$ {($M \simeq 2.3\, M_\odot$)} has a decay timescale
of about 3000 years. Hence, SURONs are not necessarily highly spinning.

{All these timescales are long enough for the initial SN to have faded
  away by the time of collapse. Moreover, these timescales are also long
  enough that the baryonic pollution of the NS surroundings will have been
  cleared out, e.g., through ejection, fall-back, or magnetic winds, such
  that a magnetosphere is established. Clear evidence for this is given
  by the Crab pulsar, which is pulsating already $10^3$ years after its
  formation.}

{We also note that neutrino cooling and the associated mass loss is
  important only during the first 10-20 \secs{} after the formation of the
  proto-neutron star (12 \secs{} were measured for SN1987A), which can be
  as large as 10\% of the mass. Only a few days later, however, the
  temperature has already decreased to $\sim 0.1\,{\rm MeV}\sim
  10^{9}\,{\rm K}$, so that even if the star cooled down to a zero
  temperature it would lose at most an equivalent mass $\sim 0.1\,{\rm
    MeV}\,\times 10^{57}\,{\rm nucleons}\, \simeq 10^{50}\,{\rm erg}\,
  \simeq 10^{-4}\,M$. In other words, while neutrino cooling can lead to
  observational signatures also thousands of years after the NS
  formation, see, e.g., \citet{PagePrakashLattimer2011a}, it soon becomes
  irrelevant in determining the equilibrium properties of the SURON.}

{Hence, we argue that, if SURONs are formed, they will have to
  collapse eventually and the collapse will take place in a clean
  environment that is rather different from that typically invoked for
  GRBs.}

The rates estimated by \citet{ThorntonStappersBailes2013a} for fast radio
bursts are $\sim$ 0.25 deg$^{-2}$ day$^{-1}$. This we can compare to the
observed rate of core collapse SN, e.g.,
\citet{DahlenStrolgerRiess2004a}, which is within a factor of two from
the core-collapse SN rate expected from the star formation rate
\citep{HoriuchiBeacomKochanek2011a}.

{Extrapolating this to $z=1$, using the fitting function provided in
  \citet{HoriuchiBeacomKochanek2011a}, we arrive at a rate of 8 SNe
  deg$^{-2}$ day$^{-1}$. Hence, we need only about 3\% of the massive
  stars to undergo a collapse as we envisage here to produce the fast
  radio bursts.} Given that the fraction of rapidly-rotating massive
stars could be up to 20\%, mainly due to binary interaction
\citep{de-MinkLangerIzzard2013a}, this suggestion does not seem
unreasonable. {Rotation also plays an important role in models for
  progenitors of long GRBs \citep{YoonLangerNorman2006a}, where an almost
  critically rotating iron core is needed. Spin-down of the progenitor,
  coupled to stellar mass loss \citep{Langer1998a}, might then introduce
  some metallicity bias, such that the SURON formation rate increases
  with lower metallicity.}

The main uncertainty, however, is how a SN with a rapidly rotating
progenitor proceeds to the NS stage. In principle, rapidly spinning
NSs with high magnetic fields can be formed \citep{FryerWarren2004a}
in a SN, but the magnetic field could also cause the NS to slow down
already during the formation process. Still, it is not inconceivable
that at least SURONS which are a few percent above critical mass and
hence only need relatively modest spins can be formed. Therefore, the
important question to address in the future is whether the rate of
SURONs formed in direct SN explosions can ineed be on the few percent
level.

Given that binary interactions play such an important role, {it
  is possible to consider also} an alternative formation channel, where a
slowly rotating NS gains mass and spin through accretion from its
companion in ultra-compact x-ray binaries
\citep{van-HaaftenNelemansVoss2013a} as in the standard scenario
\citep{Bhattacharyavan-den-Heuvel1991a} for ms-pulsars (MSPs).

In fact, the so-called ``black widow'' systems can show rather massive
NSs, with masses up to 2 $M_\odot$
\citep{DemorestPennucciRansom2010a}, which might have already been
born massive before accretion and
spin-up \citep{TaurisLangerKramer2011a}. The problem here is that
spin-up seems to suppress the magnetic field and it is not clear
whether accretion-induced spin-up can preserve the strong magnetic
fields needed for the emission.If SURONs were produced in this way, the
number of detectable ones could be small because of the fast spin-down
and short lifetime for the most massive stars. 

Lastly, {it is possible to consider} a formation scenario that
was originally proposed to explain MSPs, where a NS and a white dwarf
merge \citep{van-den-HeuvelBonsema1984a}. This too might lead to a
rapidly spinning, highly magnetized, and ``overweight'' NS. Any scenario
involving pulsar recycling rather than a direct formation, however, would
not follow the high-mass star formation rate and hence can be
observationally discriminated with more bursts to come.

\section{Emission from the collapse}\label{sec:emission}

Next we estimate the observational signatures of the collapse of a
SURON. Given that pulsar-emission mechanisms are notoriously difficult to
calculate, we concentrate on an order-of-magnitude description and a
discussion of the basic mechanisms. The relevant timescale for the
collapse is the free-fall timescale, $\tau _{\text{ff}}=
0.04\;r_{10}^{3/2}\, m_2^{-1/2}\;{\text
  ms}$~\citep{BaiottiHawkeMontero2005a,Baumgarte2003,GiacomazzoRezzollaStergioulas2011a,LehnerPalenzuelaLiebling2012a}.

Within a few $\tau _{\text{ff}}$, the crust of the NS, that is crucial
for thermal X-ray emission, will be covered by the emerging event
horizon. {For example, the time from the first appearance of the
  event horizon near the center of the NS until the NS surface is
  entirely covered by the event horizon can be as short as 0.15 ms for a
  slowly rotating star with dimensionless spin $J/M^2\sim0.2$
  \cite[][Fig.~16, left panel]{BaiottiHawkeMontero2005a}. This is a
  situation {we} could envisage for our SURONs. For a star near
  break-up and with $J/M^2\sim0.54$, this duration increases to 0.35 ms
  \cite[][Fig.~16, right
    panel]{BaiottiHawkeMontero2005a}\footnote{{Indeed, it has been
      shown that the timescale for the collapse increases quadratically
      with the dimensionless
      spin~\citep{BaiottiHawkeRezzolla2007a}.}}. This is also the
  timescale over which the largest changes in stellar structure take
  place { and most potential energy is liberated}. Hence, there is a
  marked difference with respect to binary NS mergers and short GRBs,
  whose characteristic timescales are at least one order of magnitude
  longer.  }

{Given the short timescale of the collapse, heat generated in
  the collapsing NS cannot be transported to the surface and radiated
  away efficiently, so that thermal emission from the surface will be
  very weak.}  The \MS, on the other hand, is the only part of the NS
which will not disappear in the collapse as it is well outside the
NS. According to the no-hair theorem, which prevents magnetic fields
from puncturing the event horizon, the entire magnetic field should in
principle detach and reconnect outside the horizon. This results in
large currents and intense electromagnetic emission. {Though the
  validity of the no-hair theorem in this context has been questioned
  by some \citep{LyutikovMcKinney2011a}, a strong magnetic shock wave
  moving at near the speed of light is indeed seen in 3D resistive
  magnetohydrodynamic simulations of the collapse of non-rotating NSs
  \citep[][Fig.~14]{DionysopoulouAlicPalenzuela2013a}.}

The total power that can be radiated by the \MS in the collapse is
given by $P_{\text{MS}}=\eta_{\rm B} (B^2/4 \pi)\,V/\Delta t.$ Given
that the magnetic field in the \MS is decaying quickly with radius we
here consider only a small shell comparable to the NS radius around
the star, i.e. {the volume is $V\simeq4 \pi (2R)^3/3$.}
Moreover, $\eta_{\rm B}$ is the fraction of magnetic energy that is
available for dissipation and $\Delta t= 1\,{\rm ms}\, t_{\rm ms}$ is the
observed burst length. This is an upper limit, as the observed pulse
widths could have been broadened by scattering.

\citet{DionysopoulouAlicPalenzuela2013a} have also computed the temporal
evolution of the ejected magnetic luminosity for the non-rotating
case. Their Fig.~15 shows a dominant peak of order $\sim0.1$ ms width
($\sim2\times\tau_{\text{ff}}$) after the event horizon has formed, and a
fainter precursor produced during the actual collapse.  The dominant
pulse is followed by additional pulses, which decay exponentially and
signal the ringdown of the newly formed black hole. Precursor and the two
leading pulses carry most of the transmitted power and contain about
$5\%$ of the available magnetic energy. Accordingly, we will use
{$\eta_{\rm B}=0.05\,\eta_{\rm B,5\%}$} in the following. The
bulk of the energy is released within 0.5 ms.

The available power in the magnetosphere of a typical pulsar,
{
\begin{equation}
\label{eq:powerMS}
P_{\text{MS}} \simeq 4.2 \times 10^{43} \; 
\eta_{\rm B,5\%} \, t_{\rm ms}^{-1}\, b_{12}^2\,r_{10}^{3}
\;\; {\text{erg}}{\,\text{\secs{}}^{-1}}\,,
\end{equation}
}\noindent
is thus of the right order of magnitude compared to the observations
[\cf Eq.~(\ref{eq:frbluminosity})]. 

For simplicity we describe the \MS with a simple aligned dipolar magnetic
field $B$, rotating at $f \Omega_{\rm K}$ and filled with a pair plasma
with particle number density $n_{\rm e}$, which we take to be a factor
$\kappa_{\rm GJ}$ times the Goldreich-Julian density
\citep{GoldreichJulian1969a}
{
\begin{equation}
n_{\rm e} = \kappa_{\rm  GJ}\frac{B \Omega}{2 \pi c e } 
\simeq 1.8 \times  10^{13}\;
f_{0.1}\, \kappa_{\rm  GJ}\, b_{12} \, m_2^{1/2} r_{10}^{-3/2} \;\;\text{cm}^{-3}\,.
\end{equation}
In principle $\kappa_{\rm GJ}$ can have any value, but for
  our disucssion here we will typically assume that it is of
  order unity, $\kappa_{\rm GJ} \sim O(1)$.}

In a standard pulsar \MS electron and positrons (e$^+$/e$^-$) are
spatially separated from each other in current sheets and glued to the
magnetic field lines by the strong Lorentz forces. The strong magnetic
shock waves generated in the collapse will accelerate the \emep pairs
over spatial scales $\sim R$. Because the gyro radius of any of these
particles is only $R_{\rm gyr} \simeq 1 \times 10^{-9}\,b_{12}$ cm ---
smaller than any radio wavelength, the particles will flow essentially
along the magnetic field lines. {Processes like synchrotron
  emission or self-absorption are therefore not applicable}.

To describe the emission, we therefore use a basic relativistic curvature
radiation model \citep{GunnOstriker1971a,RudermanSutherland1975a} over
radius $R$. {This is, in fact, the typical scale of the distortions
  of the magnetic field lines induced by the collapse. Curvature
  radiation simply describes how relativistic electrons radiate when
  following a bent trajectory. Hence, this is an almost unavoidable
  emission process under these conditions.}

The emitted power for a single relativistic electron or
positron with Lorentz factor $\gamma$ is then $P_{e} = 2\gamma^4
e^2c/3R^{2}$, with a characteristic frequency
{
\begin{equation}\label{eq:nucurv}
\nu _{\text{curv}} = \frac{3 c \gamma ^3}{4 \pi R}
 \simeq  {7.2 \;\gamma ^3}\,{r_{10}^{-1}}\;\;\text{kHz}\,.
\end{equation}
}\noindent
When $\gamma \sim 1$, the corresponding wavelength is comparable to the
size of the shock wave and the entire emission is
\emph{coherent}\footnote{The $N^2$ scaling of impulsive coherent emission
  mechanisms of charged particles in magnetic fields is in fact
  experimentally demonstrated by low-frequency radio emission observed
  from cosmic ray air shower fronts
  \citep{FalckeGorham2003a,FalckeApelBadea2005}, which has inspired our
  treatment.}, giving a total power of $P_{\rm t}= N_{\rm e}^2P_{\rm e}$,
where $N_{\rm e}= n_{\rm e}\,V$ is the total number of particles. At
higher frequencies and higher $\gamma$, the coherence length
{remains one wavelength but becomes smaller than the emitting region
  \citep[e.g.,][]{FalckeGorham2003a,AloisioBlasi2002a}. Hence, we have a
  number $N_{\rm slices}\simeq V\, R^{-2}
  (c/\nu_{\text{curv}})^{-1}=\gamma^3$ of coherently emitting slices
  perpendicular to the line of sight} (defined such that $N_{\rm slices}
=1$ for $\gamma=1$).

The emitted power then becomes $P_{\rm t}=\eta_{\rm e}\, N_{\rm
  slices}^{-1} \, N_{\rm e}^2\, P_{\rm e}$, where $\eta_{\rm e}$ accounts for the
fraction of \emep pairs that are accelerated to Lorentz factor
$\gamma$. Hence,
{
\begin{equation}\label{eq:totalpower}
P_{\rm t} \simeq {7.0 \times  10^{43} \;\eta_{\rm e}\, \gamma\, f^2_{0.1}\, 
\kappa_{\rm GJ}^{2}\, b_{12}^2\, m_2\,  r_{10}}
\;\;\text{erg}\,\text{\secs{}}^{-1}\,.
\end{equation}
}\noindent
Simple energy conservation imposes that $P_{\text{MS}}\ge P_{\rm t}$ and
it follows from Eqs.~(\ref{eq:powerMS}) and (\ref{eq:totalpower}) that
{
\begin{equation}
\label{eq:gammamax}
\gamma_{\rm max}\le \frac{9 c \eta _{\rm B}}{8 \, R \,\Delta t\, 
\kappa_{\rm GJ}^2 \, \eta_{\rm e}\,\Omega^2} \simeq 
0.6 \; \eta_{\rm e}^{-1}\, \eta_{\rm B,5\%}\,f^{-2}_{0.1}\,t_{\rm ms}^{-1}
\, \kappa_{\rm GJ}^{-2}\, r_{10}^{2} \,m_2^{-1},
\end{equation}
where again $\Delta t= 1\,{\rm ms}\, t_{\rm ms}$ is the duration of the pulse.}

However, for the radio emission to propagate through the plasma, the
radiation also has to be above the plasma frequency for a \emep pair
plasma, $\omega_{\rm p}=\sqrt{4\pi n_{\rm e} e^2/m_{\rm e}}$, which, {for
  $n_{\rm e} = n_{\rm GJ}(\kappa_{\rm GJ})$}, is
{
\begin{equation}
\label{eq:nuplasma}
\nu _{\text{p}}  = \frac{\omega_{\rm p}}{2\pi} = 
\sqrt{\frac{e B \Omega}{2\pi^2 c m_{\rm e}}} \simeq
38.6 \;{f}^{1/2}_{0.1} \,\kappa_{\rm GJ}^{1/2}\, {b_{12}^{1/2}}\, m_2^{1/4}\, 
{r_{10}^{-3/4}} \;\;\text{GHz}\,.
\end{equation}
}\noindent

For $\gamma\sim1$ the ensuing kHz radio emission would be absorbed in the
plasma and converted to plasma waves, which would again heat the
\emep. Hence, in order for the radiation to escape effectively we need
$\nu _{\text{curv}}\ga\nu _{\text{p}}$. Combining
Eqs.~(\ref{eq:nuplasma}) and (\ref{eq:nucurv}), we find {that in
  order to radiate the electrons need to have at least a Lorentz factor}
{
\begin{equation}
\label{eq:gammamin}
\gamma_{\rm min} \ga {175.3\, f^{1/6}_{0.1}\, \kappa_{\rm GJ}^{1/6} 
\, b_{12}^{1/6}\, m_2^{1/12}\, r_{10}^{1/12}}\,.
\end{equation}
}
To reconcile Eqs.~(\ref{eq:gammamax}) and (\ref{eq:gammamin}), and to
avoid that more power is radiated than is available in the \MS,
{we need to require} that the fraction of relativistic electrons
with $\gamma\ge\gamma_{\rm min}$ should not be higher than {
\begin{equation}
\label{eq:efficiency}
\eta_{\rm e,max} \la 0.3\,\%\; \eta_{\rm B,5\%}\, f^{-13/6}_{0.1}\, t_{\text{ms}}^{-1}
\, \kappa_{\rm GJ}^{-13/6} \, r_{10}^{23/12}\,
b_{12}^{-1/6} m_2^{-13/12}\,.
\end{equation}
}

{If we assume that the energy distribution of electrons in the
  shock front is a power-law of the form $dN_{\rm
    e}(\gamma)/d\gamma\propto\gamma^{-p}$, with $1<\gamma<\infty$, then
  $\eta_{\rm e,max} = N(\gamma \ge \gamma_{\rm
    min})/N(\gamma\ge1)=\gamma_{\rm min}^{1-p}$. Hence, after requiring
  that $\gamma_{\rm min} \ga {175}$ [\cf Eq. (\ref{eq:gammamin})] and
  that $\eta_{\rm e,max} \la 0.3\,\%$ [\cf Eq. (\ref{eq:efficiency})], we
  conclude that $p\ga2.1$, which is not uncommon in astrophysical
  sources.}

However, the very effective energy loss due to curvature radiation will
ensure that the bulk of the e$^+$/e$^-$ pairs also cannot have much
higher Lorentz factors than in Eq.~(\ref{eq:gammamin}). Hence, the
magnetosphere is already effectively dissipated by the GHz radio
emission, which is bright enough to explain the observed fast radio
bursts.

{This also naturally limits the maximum particle energy and suppresses
  additional X-ray or gamma-ray emission.} 
Of course, higher Lorentz factors can still be reached by a smaller
fraction of e$^+$/e$^-$ pairs for which $\eta_{\rm e}(\gamma)\ll1$
{as long as they are not energetically dominant.} {Given
  that the available energy in the electrons is many orders of
  magnitude less than a typical GRB energy, this emission will not be
  detectable with current detectors.}

{We point out that the spectrum of curvature radiation is
flat. However, given the rather complex nature of the shocked,
exploding \MS and the mixing between particles, plasma waves, and
radio emission the resulting spectrum could be markedly
different. This requires a more targeted effort, than what we can do
here.}

{Before concluding, a couple of remarks. First,} we have
assumed spherical symmetry, so that beaming does not play a role in
the energy budget. However, for a dipolar magnetic field the shock
wave will have a bi-polar anisotropy and the emission could be beamed
and increase the observed flux. Even for isotropic emission, the
relativistic beaming will have the effect that the observer sees only
a small patch of the emitting region. This, together with the ordered
structure of the \MS, could lead to significant polarization of the
radiation. {Moreover, absorption or scattering in the pulsar
  magnetosphere could modify the emission pattern. Obviously, radio
  emission can propagate unhampered along the polar axis but it could
  be scattered and absorped in a compact toroidal region as seen,
  e.g., in the double pulsar \citep{BretonKaspiMcLaughlin2012a}}.

{Second, our estimates assume $\kappa_{\rm GJ}\sim1$ and
  $\eta_{\rm B,5\%}=1$. The estimates of
  \citet{LehnerPalenzuelaLiebling2012a} in the case of rapidly rotating
  stars indicate that rotation in general increases the efficiency, but
  only up to $\eta_{\rm B,5\%} \simeq 4$ (i.e.~$\eta_{\rm
    B}=20\%$). Hence, even when considering the extreme case of a very
  rapidly rotating pulsar, the order-of-magnitude estimate of the power
  in Eq.~(\ref{eq:powerMS}) is robust, with changes of at most a factor
  of four with rotation. The most constraining relation, however, is
  Eq.~(\ref{eq:efficiency}), which depends inversely on a high power of
  $\kappa_{\rm GJ}$ and which could in principle vary by many orders of
  magnitude in all directions for a generic model of FRBs. However, for
  $\kappa_{\rm GJ}<7\%$ {we} would get $\eta_{\rm e,max}>1$,
  given our scaling, and the model would break down. It is thus
  intriguing to note that -- in logarithmic terms -- the minimum density
  needed to establish a normal pulsar \MS is not too far above the
  minimum density needed to explain FRBs.}

\section{Summary and Discussion}\label{sec:discussion}

{We have argued that the short time scale of the observed
  fast radio bursts, if indeed cosmological, may be indicative of a NS
  collapse. Moreover, the strong pulsar-like radio emission argues for
  emission associated with a \MS and low baryon content. Supramassive
  rotating NSs can in principle provide such a setting.}

If SURONs are formed, {and there is no reason to believe they should
  not form}, then they would collapse within several thousand to million
years due to magnetic braking. The collapse of a SURON would proceed
mostly quietly, producing a strong electromagnetic pulse due to the
strong snapping of the magnetic field in the \MS. Such a radio-emitting
collapsar, which we dub \emph{``blitzar''}\footnote{Blitz (German) =
  lightning flash} due to its bright radio flash, could be a viable
explanation for the recently discovered fast radio bursts. The parameters
needed for the NS are not significantly different from those of normal
young pulsars, except for the higher mass. {None of the processes we
  invoke in our scenario are in any way exotic}. The SURON even need not be
spinning very rapidly, as long as the mass excess is small.

Nonetheless, the energy demands to produce a fast radio burst from a
blitzar could quickly increase, if the observed radio spectrum is seen
to extend with a flat spectrum to high frequencies or if the radiative
efficiency of the \MS is even lower than we assume. {In this case,
magnetic fields in excess of $10^{12}$ G could be needed, which are
in fact observed in the pulsar population.}

Formation scenarios for SURONs could involve a direct collapse in a SN
explosion or spin-up due to accretion or merger with a white dwarf. It
remains to be seen if any of those scenarios can produce magnetized
SURONs at a sufficient rate and with high enough magnetic fields. The
formation of very short-lived SURONs could have observational
consequences in the form of an observed extended plateau phase in GRBs
\citep{LipunovaGorbovskoyBogomazov2009a,Rowlinsonetal2013}. SURONs
should also exist in our own Galaxy, but only as a small fraction of
the pulsar population due to their short lifetime and the small
fractional birthrate.

{Future searches for fast radio bursts can determine whether
  blitzars indeed trace the star formation rate in the universe and
  whether there is a connection to metallicity. This would lend
  support for an association with NS formation. Simultaneous optical
  and X-ray data could constrain the emission process, baryon load,
  and the delay of the collapse.}

{Finally, we point out that the picture sketched here could provide
  an interesting window onto the formation of isolated stellar-mass BHs,
  which would be otherwise invisible because the corresponding
  gravitational-wave emission is small. The ringdown of the event horizon
  could be visible in the radio emission of a blitzar as a succession of
  exponentially decaying sub-ms pulses. Detecting these decaying pulses
  will require observations with high signal-to-noise and low
  intergalactic and interstellar scattering, but would provide a unique
  signature for the birth of a BH. 
}

\subsubsection*{Acknowledgments} We thank {A. Achterberg, S. Buitink,
  D. Champion, S. Johnston, J. Hessels, W. Hermsen, B. Gaensler,
  M. Kramer, P. Kumar, N. Langer, P. M\'esz\'aros, T. Oosting,
  P. Pizzochero, L. Stella, L. van Haaften, and F. Verbunt} for
discussions and useful suggestions. We also thank the first referee
for timely and the second referee for constructive comments that have
improved the manuscript. HF acknowledges support from an Advanced
Grant of the European Research Council under the European Union's
Seventh Framework Program (FP/2007-2013)/ERC Grant Agreement
n. 227610. LR acknoweldges support from the DFG grant SFB/Transregio~7
and by ``NewCompStar'', COST Action MP1304.


\bibliographystyle{aa} 
\bibliography{hfrefs}

\end{document}